\def\PRL{{ Phys. Rev. Lett.\ }\/}
\def\PRB{{ Phys. Rev. B\ }\/}
\def\PRX{{ Phys. Rev. X\ }\/}

\def\etal{{\it et al.,~}\/}

\def\be{\begin {equation}}
\def\ee{\end {equation}}
\def\ber{\begin {eqnarray}}
\def\eer{\end {eqnarray}}
\def\bers{\begin {eqnarray*}}
\def\eers{\end {eqnarray*}}

\makeatletter

\newcommand{\Rmnum}[1]{\expandafter\@slowromancap\romannumeral #1@}
\makeatother

\makeatletter
\newcommand*\env@matrix[1][*\c@MaxMatrixCols c]{%
  \hskip -\arraycolsep
  \let\@ifnextchar\new@ifnextchar
  \array{#1}}
\makeatother

\documentclass[aps,prb,showpacs,superscriptaddress,a4paper,twocolumn]{revtex4-1}
\usepackage{amsmath}
\usepackage{float}
\usepackage{color}

\usepackage[normalem]{ulem}
\usepackage{soul}
\usepackage{amssymb}

\usepackage{amsfonts}
\usepackage{amssymb}
\usepackage{graphicx}
\begin {document}

\title{Unique Dirac and Triple point fermiology in simple transition metals and their binary alloys}

\author{Chiranjit Mondal}
\thanks{These two authors have contributed equally to this work}
\affiliation{Discipline of Metallurgy Engineering and Materials Science, IIT Indore, Simrol, Indore 453552, India}

\author{Chanchal K. Barman}
\thanks{These two authors have contributed equally to this work}
\affiliation{Department of Physics, Indian Institute of Technology, Bombay, Powai, Mumbai 400076, India}

\author{Shuvam Sarkar}
\affiliation{UGC-DAE Consortium for Scientific Research, Khandwa Road, Indore 452001, Madhya Pradesh, India}

\author{Sudipta Roy Barman}
\affiliation{UGC-DAE Consortium for Scientific Research, Khandwa Road, Indore 452001, Madhya Pradesh, India}

\author{Aftab Alam}
\email{aftab@iitb.ac.in}
\affiliation{Department of Physics, Indian Institute of Technology, Bombay, Powai, Mumbai 400076, India}

\author{Biswarup Pathak}
\email{biswarup@iiti.ac.in }
\affiliation{Discipline of Metallurgy Engineering and Materials Science, IIT Indore, Simrol, Indore 453552, India}
\affiliation{Discipline of Chemistry, School of Basic Sciences, IIT Indore, Simrol, Indore 453552, India}

\date{\today}

\begin{abstract}

Noble metal surfaces (Au, Ag and Cu etc.) have been extensively studied for the Shockley type surface states (SSs). Very recently, some of these Shockley SSs have been understood from the topological consideration, with the knowledge of global properties of electronic structure. In this letter, we show the existence of Dirac like excitations in the elemental noble metal Ru, Re and Os based on symmetry analysis and \textit{first principle calculations}. The unique SSs driven Fermi arcs have been investigated in details for these metals. Our calculated SSs and Fermi arcs are consistent with the previous transport and photo-emission results. We attribute these Dirac excitation mediated Fermi arc topology to be the possible reasons behind several existing transport anomalies, such as large non-saturating magneto resistance, anomalous Nernst electromotive force and its giant oscillations, magnetic breakdown etc. We further show that the Dirac like excitations in these elemental metal can further be tuned to three component Fermionic excitations, using symmetry allowed alloy mechanism.

\end{abstract}
	
\maketitle

Symmetry protected multi-fold band crossings in momentum space often exhibit strong topological response in the transport measurement. A four-fold Dirac node\cite{Dirac-1} splits into a pair of two-fold Weyl nodes\cite{Weyl-1} under magnetic field, which in turn shows several anomalous transport signatures; such as anomalous Hall effect (AHE),\cite{AHE2014,AHE2017,MirrorAnomaly,AHECo3Sn2S2} anomalous Nernst effect (ANE),\cite{PNAS2018,ANECd3As2,ANETaAsP} non-saturating large magneto-resistance (LMR),\cite{lmrPtSe2,MR2016-1,MR2016-2} chiral anomaly\cite{ChiAnomaly-1,ChiAnomaly-2,ChiAnomaly-3,ChiAnomaly-4} etc. A pair of opposite monopole charges are created upon the separation of Weyl nodes under either inversion or time reversal symmetry (TRS) breaking conditions. Each of the Weyl nodes are associated with the source or sink of Berry curvature in momentum space.\cite{Vishwanath2018,XWan2011} While this fictitious magnetic field like Berry curvature couples to the external magnetic field, it gives rise to such anomalous response in materials. Several Dirac and Weyl semi-metals (DSM and WSM) have been proposed and their topological signatures have been extensively investigated through photo-emission and transport measurements. Another type of quasi-particle excitation, different from DSM and WSM is triple point semi-metal (TPSM) states.\cite{TPMetal2016,MoP, MoC, SimpleHeusler,QHeusler, NaCu3Te2-1, NaCu3Te2-2} TPSM is believed to be an intermediate phase of relatively higher symmetric DSM and lower symmetric WSM. The topological index for TPSM is still a matter of debate,\cite{Mehdi2016,Mehdi2018} hence it has become a fertile ground for the topological study in the recent times.

In the present study, we investigated the topological electronic structures of hexagonal noble metals ruthenium (Ru), rhenium (Re) and osmium (Os) based on symmetry analysis and \textit{first principle calculations} (see sec. I of supplemental Material (SM) \cite{supp} for details of computational method). These systems although look simple on the onset, yet some of their properties are quite puzzling and still leaves proper understanding. One of the main motivation to choose these systems is to understand the rich physics behind the various anomalous existing experimental results such as anomalous magneto transport effect, \cite{Ru-1, Ru-2} anomalous Nernst emf (and their giant quantum oscillation), \cite{Ru-10} measured Fermi surface (FS),\cite{Ru-13} etc. Close inspection of these experimental results made us speculate the topological origin of the electronic structure of these systems to be responsible for such anomaly. Indeed, our detailed calculations and group theoretical analysis strongly indicate the existence of symmetry protected multiple Dirac Fermionic excitations near the Fermi level (E$_F$). We choose Ru as a case study and investigate both the bulk and surface band topology in details. Our calculated Fermi surfaces (FS) for Ru matches fairly well with the previously measured experimental FS. \cite{Ru-14}

Ru has been extensively studied for its unusual magneto-transport properties under the so called neck-lens magnetic breakdown.\cite{Ru-1, Ru-2} For instance, it shows non-saturating LMR in perpendicular magnetic field \cite{Ru-2} which is somewhat similar to these in topological semi-metals. Several theories have been proposed to address the origin of such LMR. They are$-$(i) linear band crossing \cite{Ru-3, Ru-4, Ru-5} in momentum space as in the case of Cd$_3$As$_2$, Na$_3$Bi and so on, (ii) perfect electron-hole (e-h) compensation \cite{WTe2-Li,type-II2018,Ru-6, Ru-7} in WSMs; WTe$_2$, MoTe$_2$, PtSn$_4$ and LaSb (although LaSb has trivial band ordering, multiple Weyl type nodes are present in its band dispersion \cite{Ru-8}), and (iii) Lipschitz transition (LT) of Fermi surface (FS) \cite{Ru-7} (recently, LT has been found in several topological materials where the phase transition does not break any symmetries but can be described by topological invariants). Apart from aforementioned theories, a topology driven non-trivial origin for the non-saturating LMR has also been predicted by Tafti et al.\cite{Ru-9} Another interesting feature of Ru is that it shows finite Nernst emf which shows giant oscillation in high filed regime. \cite{Ru-10} The characteristic curve and the oscillation patterns are quite similar to the non-trivial material Bi$_2$Se$_3$ and very different from the Drude like behavior.\cite{Ru-10, Ru-11} Keeping these facts in mind, our study reveals the appearance of Dirac surface states (SSs) mediated Fermi arcs on the surface of noble metals (Ru, Re and Os) which is found to be the key origin in understanding the existing problem of transport anomalies. Note that the controversial Shockley type SSs that appear on the surface of noble metals; such as gold (Au), silver (Ag), copper (Cu), platinum (Pt) and palladium (Pd), have recently been interpreted as topologically derived surface states. \cite{Ru-12}

Symmetry argument has become a very important tool to explore the material physics in recent time. The knowledge of symmetry provides constraint or consent over the tunability of topological phases. We use the crystalline symmetry breaking argument to tune the DSM phase to TPSM phase using alloy mechanism. For example, both Ru and Os show Dirac nature owing to the center of inversion (IS) and C$_6$ rotation symmetry. However, the binary alloy RuOs breaks the IS and transforms the C$_6$ into C$_3$ rotation, which converts the Dirac like excitations to three component Fermionic excitations. The most important advantages of such symmetry adopted tunability is that we can shift the nodal points very close E$_F$ depending on the crystal composition. As such, we have a freedom on the position of the nodal points as well as tunability of the low energy excitations.

\begin{figure}[t]
\centering
\includegraphics[width=\linewidth]{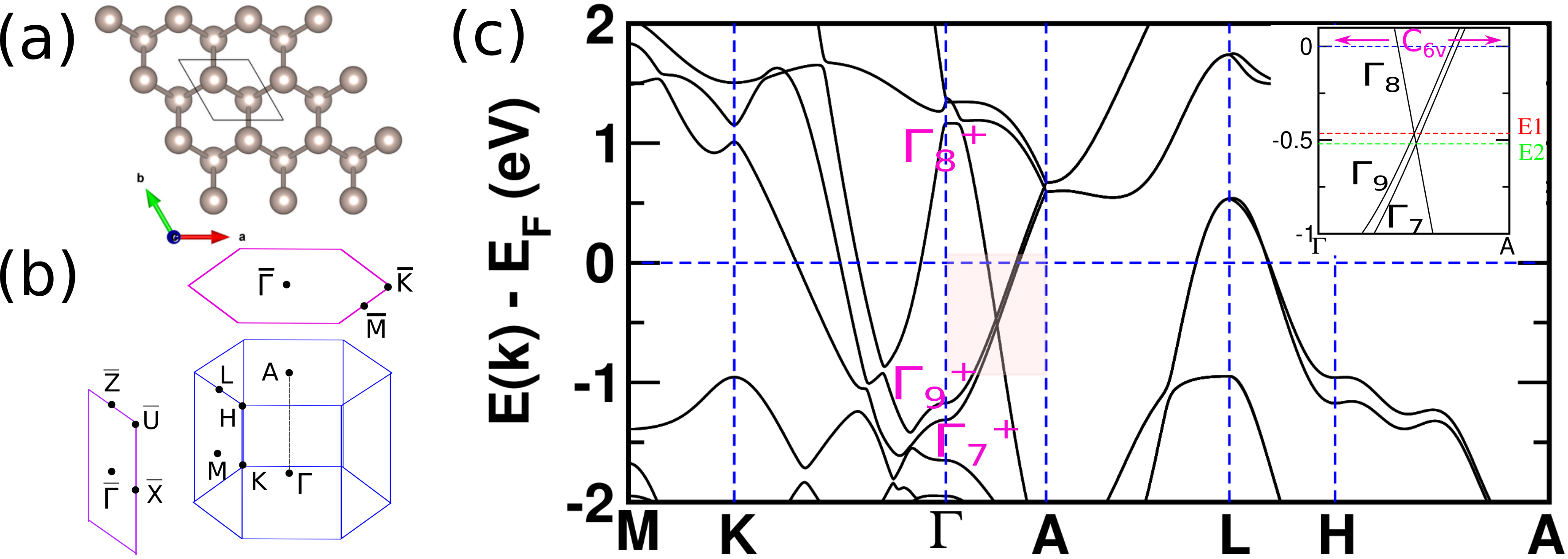}
\caption{(Color online) (a) Top view of P6$_3$/mmc crystal structure of metals (b) Brillouin zone (BZ) with high symmetry points and their projections on (001) and (100) surface BZ. (c) Electronic structure of Ruthenium (Ru). Inset in (c) shows small energy window of the shaded region along $\Gamma$-A. $\Gamma_i$'s are the irreducible representations (IRs) for the bands. The intersection of $\Gamma_{7or8}$ with $\Gamma_9$ results in Dirac nodes. Red(green) dotted line is drawn to show Dirac nodes at energy E1(E2). }
\label{fig1}
\end{figure}



Ru, Re and Os, in their elemental phase, crystallize in hexagonal-close-packed structure with P6$_3/$mmc space group and D$_{6h}$ point group. The crystal structure and Brillouin zones (BZ) are shown in Fig.~\ref{fig1}(a,b). The two-atom unit cell with particular uniaxial rotational symmetry results more complex electronic Fermi surface topology than the cubic noble metals (Au, Ag, Pt, etc). The presence of D$_{6h}$ point group allow a C$_{6v}$ little group along $\Gamma$-A direction in BZ. The symmetry elements that C$_{6v}$ contain are identity (E) operation, six (C$_6$), three (C$_3$) \& two-fold (C$_2$) rotational symmetry about z-axis, three vertical mirror plane ($\sigma_v$) and three $\sigma_d$ mirror plane ($\sigma_d$ bisect two $\sigma_v$ mirror). Under spin-orbit coupling scheme, $\tilde{C_6}$ possess six eigenvalues, namely, $e^{\pm{i\frac{\pi}{6}}}$, $e^{\pm{i\frac{\pi}{3}}}$, $e^{i\frac{\pi}{2}}$, $e^{i{\pi}}$. The corresponding eigen states for the $\tilde{C_6}$ rotation operator can be denoted as $\psi_1$, $\psi_2$, $\psi_3$, $\psi_4$, $\psi_5$, and $\psi_6$ (see  Fig. S1 of SM \cite{supp}). Now, the $\tilde\sigma_v$ (x-axis mirror) and $\tilde{C_6}$ do not commute and thus cannot be simultaneously diagonalizable in the eigen space of $\tilde{C_6}$ operator. The mirror $\tilde\sigma_v$ keeps the $\psi_5$, and $\psi_6$ invariant, i.e, $\tilde\sigma_v$ $\psi_5$ = $\psi_5$ and $\tilde\sigma_v$ $\psi_6$ = $\psi_6$ whereas $\psi_1$ will convert to $\psi_2$ and $\psi_3$ will convert to $\psi_4$ under the action of $\tilde\sigma_v$, i.e, $\tilde\sigma_v$ $\psi_1$ = $\psi_2$ and $\tilde\sigma_v$ $\psi_3$ = $\psi_4$. As the $\Gamma$-A direction is invariant under these symmetries, the non-commutating condition of $\tilde\sigma_v$ and $\tilde{C_6}$ enforces to form doubly degenerate eigen space by $\psi_1$+$\psi_2$ ($\Gamma_7$ representation designated by $e^{\pm{i\frac{\pi}{6}}}$ eigenvalues) and $\psi_3$+$\psi_4$ ($\Gamma_8$ representation designated by $e^{\pm{i\frac{\pi}{3}}}$ eigenvalues) along $\Gamma$-A. Furthermore, $\psi_5$ and $\psi_6$ form a degenerate state ($\Gamma_9$ representation designated by $e^{i\frac{\pi}{2}}$ and $e^{i\pi}$ eigenvalues) under the action of time reversal symmetry (TRS), C$_2$ and $\sigma_d$ symmetry. As $\tilde{C_2}$ and $\tilde\sigma_d$ commute, we can define a new operator, $\theta$=$\tilde{C_2}$  $\tilde\sigma_d$. In spin rotation space, $\theta^2=1$. In the presence of TRS, eventually now we have, $T^{2}\theta^2=-1$. This is the local Kramer's theorem, which guaranteed the double degeneracy at every point along $\Gamma$-A direction in BZ as this direction is invariant under both $\tilde{C_2}$ and $\tilde\sigma_d$. Now, the two bands having different irreducible representations cross each other along C$_6$ axis will form a gapless four fold degenerate Dirac nodes. For Ru, $\Gamma_{7or9}$ band intersects with $\Gamma_8$ band and form two Dirac nodes on the C$_6$ rotation axis as shown in Fig.~\ref{fig1}(c). Similar observations and mechanisms have also been observed for Re and Os which will be discussed later in the manuscript. Furthermore, in addition to the above C$_{6v}$ symmetry element, the elements Ru, Re and Os also holds structural inversion. Interplay of inversion symmetry and TRS further ensures Kramer's double degeneracy throughout the BZ. Hence, two doubly degenerate bands belong to different IRs while crossing each other along C$_6$ direction form a Dirac node and hence the hybridization at the nodal point is prohibited by group orthogonality relations. As such, presence of inversion center provides extra crystalline symmetry protection to the Dirac nodes in addition to C$_{6v}$. Therefore, the Dirac nodes are stable against inversion breaking perturbation in the presence of C$_{6v}$ symmetry. 

\begin{figure*}[t]
\centering
\includegraphics[width=\textwidth]{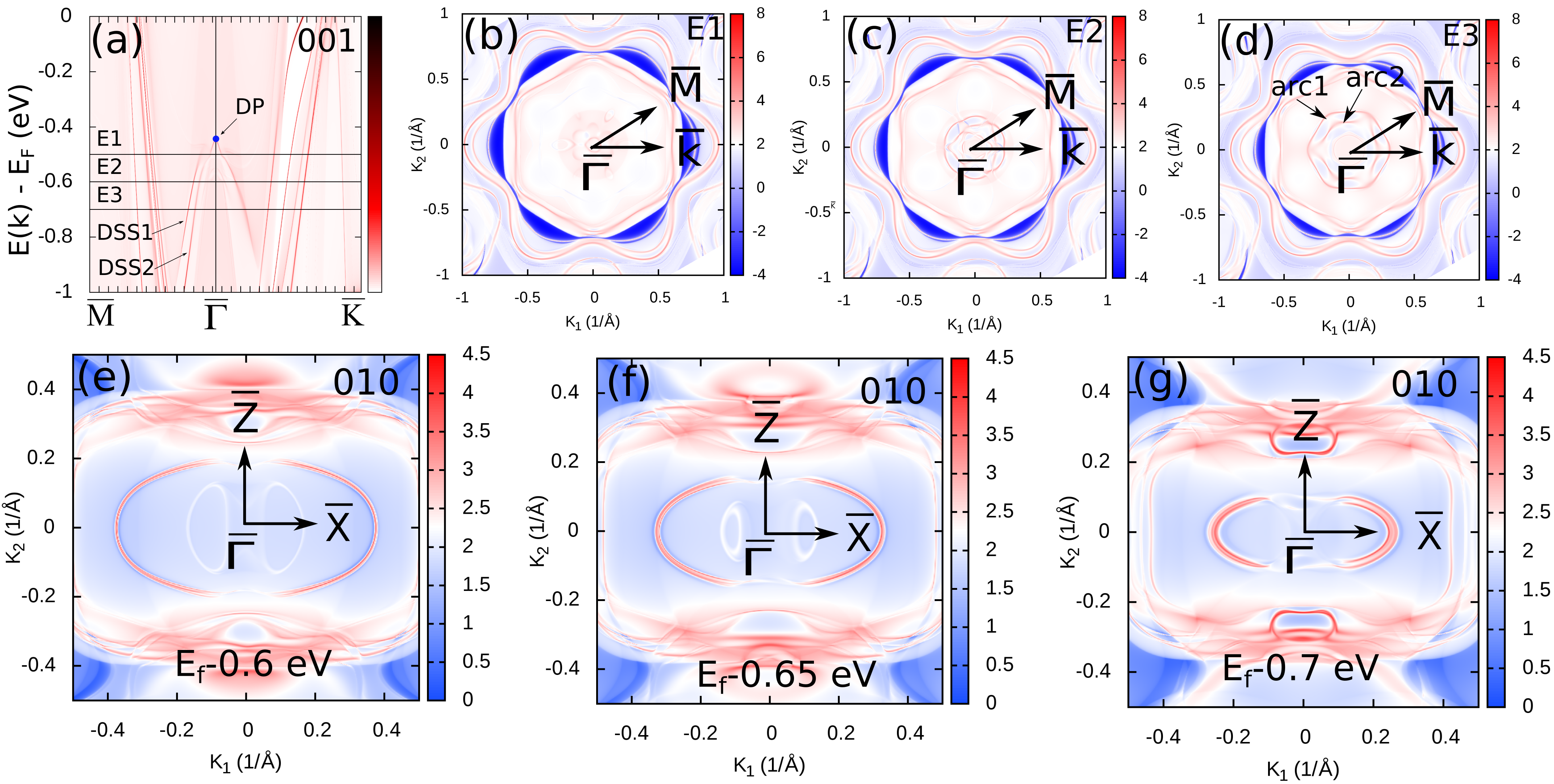}
\caption{(Color online) For Ruthenium, (a) Surface dispersion for (001) miller plane. (b-d) Fermi surface maps on (001) surface at different energy cuts. (e-g) Fermi arcs at different energy cuts on the (010) side surface.}
\label{fig2}
\end{figure*}

 The origin of the surface states (SSs) in noble metals (Au, Ag and Cu etc.) can be traced back to Shockley's prediction \cite{ss-1} of SSs which appear inside an inverted energy gap due to band crossing. \cite{ss-2, ss-3} Very recently, some of these Shockley's SSs have been understood from the topological consideration, with the knowledge of global properties of electronic structure.\cite{Ru-12} Another type of SSs, called Rashba SSs, appear due to the absence of translational symmetry on the surface which is quiet a common phenomenon on noble metal surfaces. It is to be noted that these Rashba SSs are unlike the conventional Rashba states which arises due to the breaking of inversion symmetry (in the presence of spin-orbit coupling) in bulk systems. Although, these Shockley or Rashba SSs can be explained from the free electron theory, but a rigorous topological understanding is required to capture various other anomalous surface behaviors. For example, recently, Zak phase driven large surface polarization charge and flat SSs have been understood from the non-trivial Berry phase in topological Dirac nodal line fcc alkaline earth metals; calcium (Ca), strontium (Sr) and ytterbium (Yb). \cite{ss-4} Topological nature of Beryllium (Be) have also been explored which indeed shed light on the long standing controversial issues of Be, such as strong deviations from the description of the nearly free-electron theory, anomalously large electron-phonon coupling effect, \cite{Be} and large Friedel oscillations etc.\cite{ss-5}

In the above context, Ru, Re and Os are unique as their Dirac like bulk band crossing suggests the appearance of non-trivial surface dispersion and Fermi arc topology onto the surface. Since in DSMs, the main focus is the surface states and the associated Fermi arcs that link the Dirac points, we have investigated the FS map for both (001) and (010) surfaces for Ru as shown in Fig.~\ref{fig2}. Bulk projected SSs on (001) surface is shown in Fig.~\ref{fig2}(a). Bulk electronic structure of Ru suggest the appearance of a pair of Dirac nodes (crossing point of $\Gamma_{7or9}$, and $\Gamma_8$ bands as shown in Fig.~\ref{fig1}(c)) along $\Gamma$-A direction). The projection of these two Dirac nodes on (001) surface fall on $\overline{\Gamma}$ point in the surface BZ (as shown in Fig.~\ref{fig1}(b)). Moreover, in the energy scale, the positions of two Dirac nodes are separated by a small value. Projection of one Dirac node in energy scale is depicted by blue dot in Fig.~\ref{fig2}(a). The Dirac like surface states (DSSs), emerging from the Dirac nodes in (001) surface is shown in Fig.~\ref{fig2}(a). However, the other Dirac node is buried by the bulk Fermi pockets near the $\overline{\Gamma}$ point. Nonetheless, the Fermi arcs for two Dirac nodes can be clearly encapsulated little far from the $\overline{\Gamma}$ point along $\overline{M}$($\overline{K}$) direction (see Fig.~\ref{fig2}(a)) as indicated by DSS1 and DSS2.

The FS maps on (001) surface have been investigated and the evolution of FS topology is observed for three different energy cuts (E1, E2 and E3) as shown in Fig.~\ref{fig2}(b-d). At energy cut E1, the Dirac SSs, DSS1 and DSS2 are little immersed by bulk bands near $\overline{\Gamma}$, hence we do not observe the clear signature of Dirac SSs mediated FS map at this constant energy cut. However, the signatures and contour patterns of DSSs mediated FS maps are clearly observed for other two energy cuts at E2 and E3 as described in Fig.~\ref{fig2}(c,d). The two concentric Fermi arcs in Fig.~\ref{fig2}(d) are indicated by arc1 and arc2. A similar pattern for FS have earlier been observed by photo-emission spectroscopy on the hexagonal surface of MoP, \cite{MoP} MoC, \cite{MoC} LuPtBi, \cite{LuPtBi} etc. Most importantly, the FS for Ru is highly consistent with the previous experiments both by Angle-resolved photo-emission spectroscopy (ARPES) and de Haas$-$van Alphen (dHV) oscillation. \cite{Ru-13, Ru-14} However, the explanation based on topological origin of the FS maps in those experiments is lacking.

Furthermore, the DSSs mediated Fermi arc topology, can also be observed on the side surface (010 plane) of Ru. The $\Gamma$ and A(L) point of bulk BZ fall onto the $\overline{\Gamma}$ and $\overline{Z}$ point on surface BZ (SBZ) of (010) surface. As such, the projected Dirac nodes fall on the $\overline{\Gamma}$- $\overline{Z}$ line segment in (010) SBZ. A time reversal pair of Dirac nodes is situated on both sides of the $\overline{\Gamma}$ point along $\overline{\Gamma}$- $\overline{Z}$ line. In the momentum space, a Fermi arc is nested in between this pair of Dirac nodes as shown in Fig.~\ref{fig2}(e-g). Figure~\ref{fig2}(e-g) shows the Fermi arc topology on the side surface of Ru for three different energy cuts. For the energy cut very close to Dirac nodes (as shown in Fig.~\ref{fig2}(e)), two arcs originated from two pair of Dirac nodes are almost degenerate (since, two nodal points are situated very closely in momentum space). As we move away from the Dirac nodes, the two arcs get resolved (Fig.~\ref{fig2}(f),~\ref{fig2}(g)). We have discussed and compared, in detail, our calculated Fermi arcs with the previously measured ARPES and transport results in section III of SM \cite{supp}. The close similarity of calculated Fermi arc topology (in both (001) and (010) surfaces) with the measured data validate our theoretical predictions.

\begin{figure}[t]
\centering
\includegraphics[width=\linewidth]{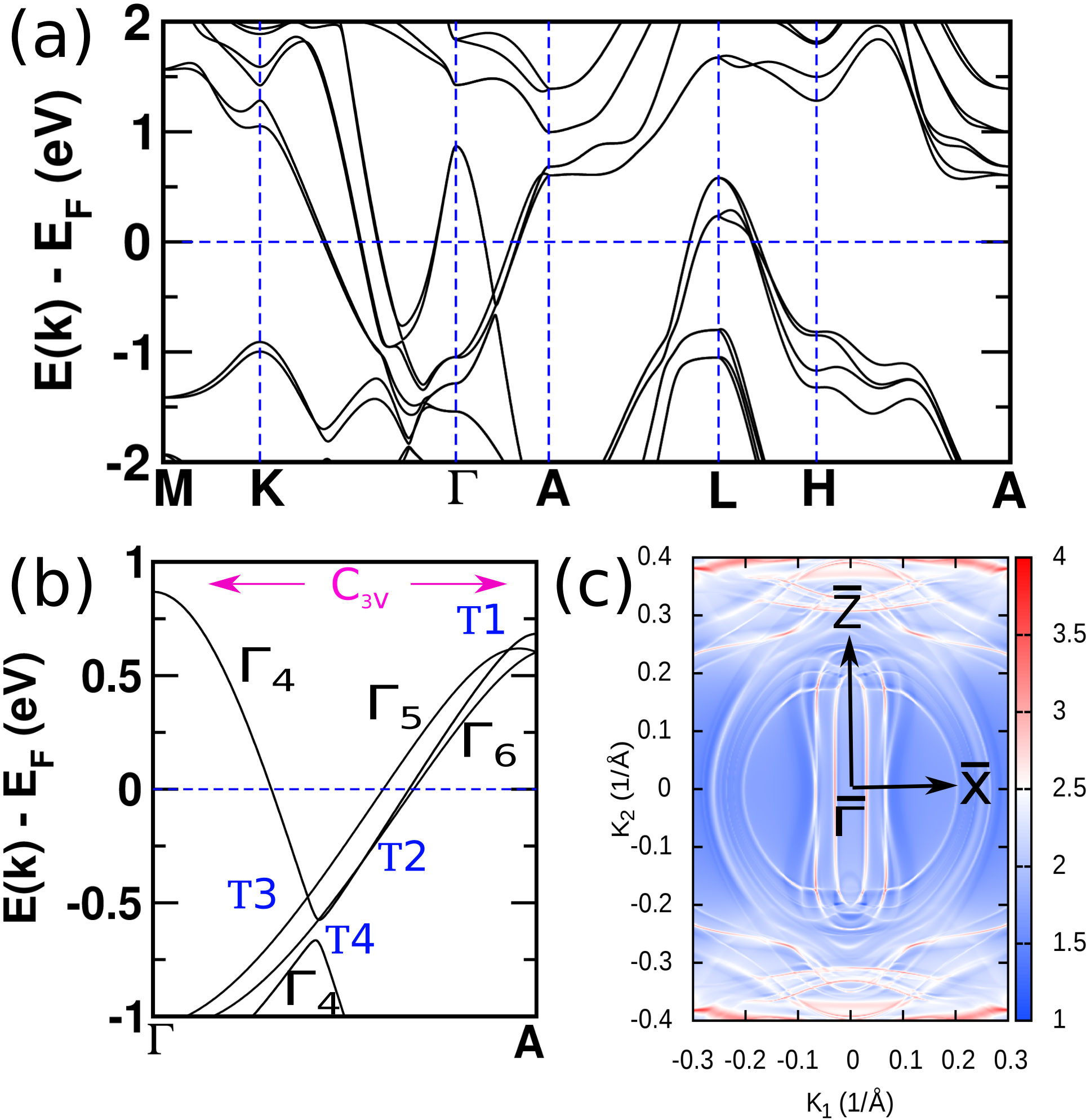}
\caption{(Color online) For RuOs binary alloy, bulk band structure (a) along high symmetry lines, (b)along $\Gamma$-A indicating band IRs (shown by $\Gamma_i$s). Triply degenerate nodal points TDNPs (shown by T's). (c) Fermi arc on (010) side surface.}
\label{fig3}
\end{figure}

We now discuss the origin of these SSs and arcs from the topological perspective. A DSM phase is the parent sate of WSM and topological insulator (TI). TI phase can be achieved by opening up a non-trivial gap at the nodal points. In such a case, presence of SSs is guaranteed by topological Z$_2$ index. On the other hand, two Weyl nodes with opposite chern number sit together to form a Dirac node in momentum space under the precise symmetry enforcement. Such degeneracies of Weyl nodes form a "doubly-degenerate" Fermi arc in DSM phase. However, such type of Fermi arcs may not be protected by topological index. Nevertheless, a crystalline symmetry protected three dimensional DSM phase is stable as long as the symmetries are intact.

Further, the alloy driven crystalline symmetry breaking allows us to realize three component Fermionic excitation near E$_F$, which is different from the Dirac excitation in pure metals Ru, OS and Re. For the binary alloys, e.g. RuOs, the point group symmetry reduces from D$_{6h}$ to D$_{3h}$. D$_{3h}$ allows it's C$_{3v}$ subgroup symmetry along $\Gamma$-A direction. The symmetry elements that C$_{3v}$ contain are E, C$_3$ and three $\sigma_v$ (see Fig. S3 of SM \cite{supp}). Similar to earlier explained C$_{6v}$ case, the non-commutation relation of $\tilde{C}_3$ and $\tilde{\sigma_v}$ (say, x-axis mirror) allows two singly degenerate states (denoted by $\Gamma_5$ and  $\Gamma_6$) and one doubly degenerate state (denoted by $\Gamma_4$) along $\Gamma$-A direction in spin-orbit space. The operation of C$_3$ and $\sigma_v$ does not alter momentum co-ordinate along \textit{k$_z$}-axis. Any accidental band crossing of $\Gamma_{5or6}$ with $\Gamma_4$ along \textit{k$_z$}-axis forms a triply degenerate nodal point (TDNP). In particular, such an alloying, transforms the crystalline symmetry from C$_{6v}$ (elemental metal) to C$_{3v}$ (binary alloy with space group P$\bar{6}$m2) and the corresponding band representation changes as; $\Gamma_{8,7}$ $\rightarrow$ $\Gamma_4$ and $\Gamma_9 \rightarrow \Gamma_5 \oplus \Gamma_6$. Further, the strength of $\Gamma_{5,6}$ band splitting and the slop of $\Gamma_4$ band collectively determine the number of TDNP (two or four in our binary alloys) on the C$_3$ rotation axis. Figure~\ref{fig3}(a,b) shows a case study on the bulk band structure of RuOs alloy. Four triple points are observed in RuOs alloy and they are denoted by T1, T2, T3 and T4 in Fig.~\ref{fig3}(b). Note that, TDNPs are protected by group orthogonality relations of different IRs. We have also simulated the Fermi arcs nesting on (010) rectangular surface of RuOs alloy, which host triple point (semi)metallic state (see Fig.~\ref{fig3}(c)). All the TDNPs are projected along the $\overline{\Gamma}-\overline{Z}$ axis on both sides of $\overline{\Gamma}$ point. The Fermi arcs are originated and nested between the TDNPs as shown in Fig.~\ref{fig3}(c). The existence of such TDNP induced Fermi arcs on a particular surface is a hallmark of TPSM state for their experimental detection.

\begin{figure}[t]
\centering
\includegraphics[width=\linewidth]{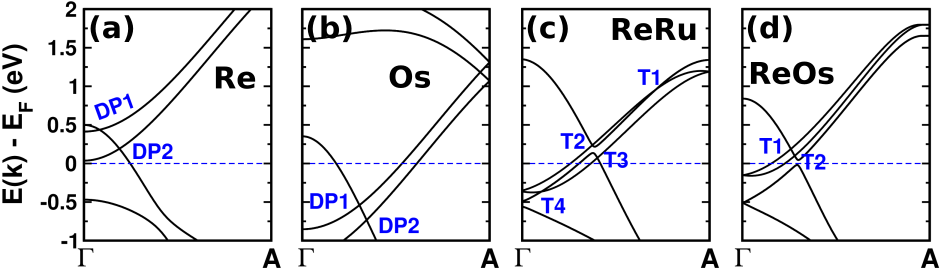}
\caption{(Color online) Electronic structure (along $\Gamma$-A) for (a) Re, (b) Os, (c) ReRu, and (d) ReOs. Dirac points and TDNPs are represented by DP's and T's respectively.}
\label{fig4}
\end{figure}

\begin{table}[t]
\begin{ruledtabular}
\caption{Number (\#) of DPs or TDNPs and their positions ($\Delta \epsilon $) with respect to E$_F$ for pure metals and their binary alloy. }
\label{Table1}
\begin{tabular}{c c c c| c c c c}
& Metal & DPs \# & $\Delta \epsilon$ (eV) & Alloy & TPs \# & $\Delta \epsilon$ (eV) \\
\hline
& Ru & 2 & -0.45   & RuOs & 4 & 0.60  \\
&    &   & -0.51   &       &   & -0.29  \\
&    &   &         &       &   & -0.47  \\
&    &   &         &       &   & -0.57   \\

& Os & 2 & -0.50   & ReOs & 2 & 0.17  \\
&    &   & -0.71   &       &   & 0.09  \\

& Re & 2 & 0.41  & RuRe & 4 & 0.94   \\
&    &   & 0.17  &       &   & 0.23   \\
&    &   &    &       &   & 0.06   \\
&    &   &    &       &   & -0.35  \\

\end{tabular}
\end{ruledtabular}
\end{table}


To get the ideal candidates (in terms of the position of Dirac points (DPs) and TDNPs with respect to E$_F$), we have simulated the band structure of other metals (Os and Re) and their alloys (ReRu and ReOs). The band structures along the six(three) fold rotation axis are shown in Fig.~\ref{fig4}. In Table~\ref{Table1}, we have also listed the compounds with the number of nodal points and their relative position in terms of energy. The spin-orbit coupling strength of Os is highest among these three pure metals, hence the splitting between $\Gamma_7$ and $\Gamma_9$ is largest which in turn results in a larger separation of DPs for Os in momentum space as shown in Fig.~\ref{fig4}(b). For ReOs alloy, the nodal points (TDNPs) lie close to E$_F$ as compared to other two alloys, ReRu and RuOs. Furthermore, ReOs only has a single pair of TDNPs, as shown in Fig.~\ref{fig4}(d). 

In order to investigate the possibilities of the experimental synthesis for these binary alloys, we have
checked the chemical and mechanical stability of
these compounds. Chemical stability is checked by calculating
the formation energies using the following formula, 
\[
 \Delta E_{form} = E_{Comp} - \sum_{i=1}^n x_i E_i
\] 
 where $E_{Comp}$ is the total energy of the binary alloy, $E_i$ represents the energy of the constituent elements in their ground state phase, and $x_i$ is the proportion of $i^{th}$ element in the binary compound. The formation energy $\Delta E_{form} $ of these three binary alloys is presented in Table~\ref{TableS3}. The negative value of $\Delta E_{form} $ confirms the chemical stability of these alloys.

\begin{table}[b]
\caption{ Formation energies ($\Delta E_{form}$) of the three binary alloys. }
\label{TableS3}
\begin{tabular}{cc|cccc} 
\hline \hline
& System\ \ \ \ \ \ \ \ \ \ \ \ \ \ & &  $\Delta E_{form}$  (meV/atom) &\\ \hline
& ReOs\ \ \ \ \ \ \ \ \ \ \ \ \ \  & & -192 & \\
& ReRu\ \ \ \ \ \ \ \ \ \ \ \ \ \  & & -176 & \\
& RuOs\ \ \ \ \ \ \ \ \ \ \ \ \ \  & & -23  & \\ \hline \hline
\end{tabular}
\end{table}

Apart from chemical stability, the elastic behavior of a lattice is described by its second-order elastic constant tensor given by,
\begin{equation}
C_{ij}=\dfrac{1}{v_0}\left(\frac{\partial^2 E}{\partial \epsilon_i \partial \epsilon_j}\right),\nonumber
\end{equation}
where E is the total energy of the crystal, $v_0$ is equilibrium volume, and $\epsilon$ denotes the strain. The stiffness tensor  is symmetric and has a size of 6$\times$6 dimension for the present alloy system. For a hexagonal system, there exists five independent elastic constants. The necessary and sufficient conditions for the elastic stability of a hexagonal lattice are,

\begin{eqnarray}
C_{11}>|C_{12}|\nonumber\\
2C_{12}^2<C_{33}(C_{11}+C_{12})\nonumber\\
C_{44}>0\nonumber\\
C_{66}>0\nonumber\\
\label{eq1}
\end{eqnarray}

The condition on $C_{66}$ is redundant for hexagonal systems with the $C_{11}$ and $C_{12}$ given by the equation, $C_{66}=(C_{11}-C_{12})/2$. 

The values of the elastic constants for the three binary alloys (Re-Os, Re-Ru and Ru-Os) are listed in Table \ref{TableS2}.  These values of elastic constants clearly satisfy the conditions, Eq. \ref{eq1}, and hence confirms the mechanical stability of the three alloys.

\begin{table}[t]
\caption{ Elastic constants of binary alloys in kBar.}
\label{TableS2}
\begin{tabular}{c c c c c c c c c c c c c c c} 
\hline \hline
& System\ \ \ \ \ \ \ & &  $C_{11}$ & & $C_{12}$ & & $C_{13}$ & & $C_{44}$ & & $C_{66}$ \\ \hline 
& ReOs\ \ \ \ \ \ \ & & 6678.8 & & 2633.1 & & 2185.9 & & 2422.7 & & 2022.9  & \\
& ReRu\ \ \ \ \ \ \ & & 5617.9 & & 2731.8 & & 1981.2 & & 1996.6 & & 1443.0  & \\
& RuOs\ \ \ \ \ \ \ & & 6542.1 & & 2169.1 & & 1949.3 & & 2165.9 & & 2186.5  & \\ \hline \hline
\end{tabular}
\end{table}

The conclusion of this work is mainly three folds; (i) we predict the existence of symmetry protected Dirac sates in pure elemental metals Ru, Os and Re. We find the unique Dirac like Fermi arc topology on the (001) and (010) surfaces of these metals. Our calculated Fermi surfaces are consistent with the previous experiments by ARPES and transport measurements. (ii) The presence of such topologically non-trivial Fermi arcs can re-evaluate the understanding behind several anomaly in such metals. We speculate these topological nature of bands in Ru to be responsible for several puzzling behavior such as magnetic break down, large magneto-resistance similar to DSM compound Cd$_3$As$_2$, giant Nernst oscillation (similar to Bi$_2$Se$_3$) and deviations from the description of the Drude theory. (iii) By precise symmetry breaking alloy engineering, the Dirac excitations can be tuned to three component fermion excitations. Depending on the combinations, we get two or four pairs of TDNPs along $\Gamma$-A directions. The position of TDNPs are very closer to E$_F$ (for RuRe and ReOs alloy) which definitely enable the strong topological response in transport experiments. Finally, these types of transition metal alloys are extensively synthesized and investigated in the field of catalysis, hence our findings open up a new direction towards that too.

{\par} This work is financially supported by DST SERB (EMR/2015/002057), India. We thank IIT Indore and IIT Bombay for the lab and computing facilities. CKB and CM acknowledge MHRD-India for financial support. AA acknowledges the National Center for Photovoltaic Research and Education (NCPRE) (financially supported by Ministry of New Renewable Energy, MNRE, Government of India) for support of this research.



\end{document}



\title{Supplemental Material \\ Unique Dirac and Triple point fermiology in simple transition metals and their binary alloys}

\author{Chiranjit Mondal}
\thanks{These two authors have contributed equally to this work}
\affiliation{Discipline of Metallurgy Engineering and Materials Science, IIT Indore, Simrol, Indore 453552, India}

\author{Chanchal K. Barman}
\thanks{These two authors have contributed equally to this work}
\affiliation{Department of Physics, Indian Institute of Technology, Bombay, Powai, Mumbai 400076, India}

\author{Shuvam Sarkar}
\affiliation{UGC-DAE Consortium for Scientific Research, Khandwa Road, Indore 452001, Madhya Pradesh, India}

\author{Sudipta Roy Barman}
\affiliation{UGC-DAE Consortium for Scientific Research, Khandwa Road, Indore 452001, Madhya Pradesh, India}

\author{Aftab Alam}
\email{aftab@iitb.ac.in}
\affiliation{Department of Physics, Indian Institute of Technology, Bombay, Powai, Mumbai 400076, India}

\author{Biswarup Pathak}
\email{biswarup@iiti.ac.in }
\affiliation{Discipline of Metallurgy Engineering and Materials Science, IIT Indore, Simrol, Indore 453552, India}
\affiliation{Discipline of Chemistry, School of Basic Sciences, IIT Indore, Simrol, Indore 453552, India}

\date{\today}

\maketitle 

\beginsupplement

\textbf{Contents}

\textbf{I. Computational details}

\textbf{II. Formation of Dirac states under C$_{6v}$} 

\textbf{III. Comparison of the simulated Fermi arcs with previous ARPES results}

\textbf{IV. Formation of triple point fermion states under C$_{3v}$}

\section{Computational Details}
First principle calculations were carried out using projector augmented wave (PAW) formalism based on Density Functional Theory (DFT) as implemented in the Vienna Ab Initio Simulation Package (VASP).\cite{1,2,3} The generalized-gradient approximation by Perdew-Burke- Ernzerhof (PBE)\cite{1} was employed to describe the exchange and correlation. Force (energy) criterion was set upto 0.01 eV/\AA(10$^{-6}$ eV). An energy cut off of 500 eV is used to truncate the plane-wave basis sets. The Brillouin zone (BZ) is integrated using 12$\times$12$\times$6 gamma centered k-mesh. Tight-binding (TB) Hamiltonians were constructed using maximally localized Wannier functions (MLWFs)\cite{4,5,6} obtained from wannier90 package. The topological properties including surface spectrum and Fermi surface were analyzed based on the iterative Green's function method\cite{7,8,9,10} implemented in Wannier-Tools package.\cite{10} 
  
\section{Formation of Dirac states under C$_{6v}$}

\begin{figure}[h]
\centering
\includegraphics[scale=0.5]{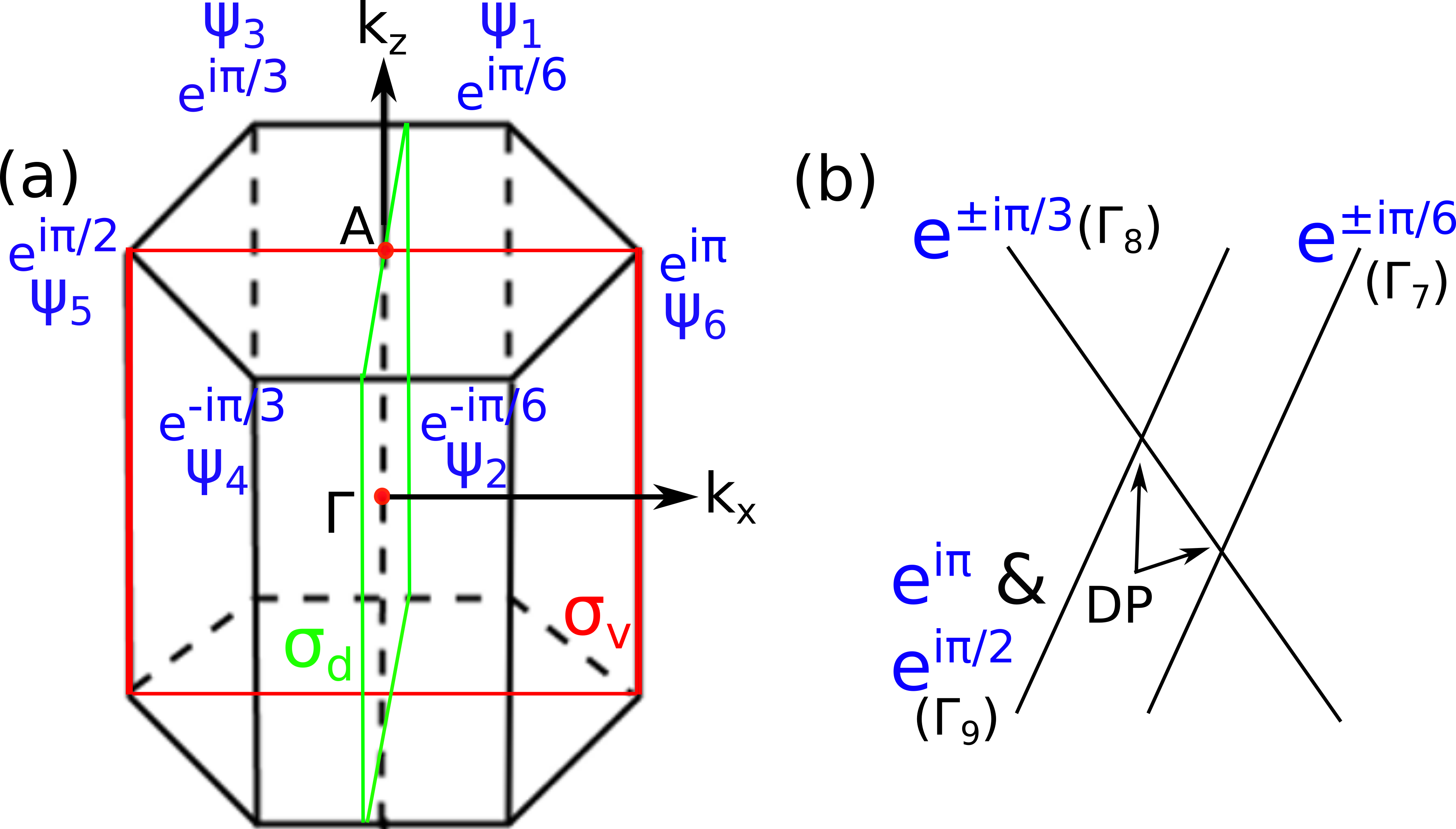}
\caption{(Color online) (a) for C$_{6v}$, representation of rotation eigen states corresponding to their rotation eigen values, and and mirror invariant planes (b) schematic diagram of formation of four-fold band crossings. Bands are represented by their rotation eigen values.}
\label{SM-fig1}
\end{figure}

The presence of D$_{6h}$ point group allow a C$_{6v}$ little group along $\Gamma$-A direction in Brillouin zone (BZ). The symmetry elements that C$_{6v}$ contain are identity (E) operation, six (C$_6$), three (C$_3$) \& two-fold (C$_2$) rotational symmetry about z-axis, three vertical mirror plane ($\sigma_v$) and three $\sigma_d$ mirror plane ($\sigma_d$ bisect two $\sigma_v$ mirror). In spin-orbit space, $\tilde{C_6}$ possess six eigenvalues, namely, $e^{\pm{i\frac{\pi}{6}}}$, $e^{\pm{i\frac{\pi}{3}}}$, $e^{i\frac{\pi}{2}}$, $e^{i{\pi}}$. The corresponding eigenstates for the $\tilde{C_6}$ rotation operator, are $\psi_1$, $\psi_2$, $\psi_3$, $\psi_4$, $\psi_5$, and $\psi_6$ as shown in Fig.~\ref{SM-fig1}(a). Now, we have the commutation relations, [$\tilde\sigma_v$, $\tilde{C_6}$]$\ne$0 and [$\tilde\sigma_v$, $\tilde{C_3}$]$\ne$0. These two commutation relation gives the following relations,

\begin{eqnarray}
\tilde\sigma_v \psi_1 &=& \psi_2  \label{eq1} \\
\tilde\sigma_v\psi_3  &=& \psi_4 \\ \nonumber
\end{eqnarray}

Now, Hamiltonian ($H$) of the system commutes with the mirror $\tilde\sigma_v$ i.e, $\left[H, \tilde\sigma_v\right]= 0$ and $H$ operating on $\psi_1 (\psi_2)$ gives eigenvalue $E_1 (E_2)$. Further, operation of the commutator on eigenstate $\psi_1$ follows;

\begin{align*}
H\tilde\sigma_v\psi_1 - \tilde\sigma_v H \psi_1 =\;& 0 \\ 
\Rightarrow H\psi_2 - E_{1}\tilde{\sigma_v}\psi_{1} =\;& 0  \quad \left(using \quad equation~\ref{eq1}\right) \\ 
\Rightarrow E_2\psi_2 - E_{1}\tilde{\sigma_v}\psi_{1} =\;& 0 \\
\Rightarrow E_2\psi_2 - E_{1}\psi_{2} =\;& 0 
\end{align*}

Thus, $E_1 = E_2$. So, the energy eigen value corresponding to $\tilde{C_6}$ rotation eigen states ($\psi_1$ and $\psi_2$) are same. Hence, they form a degenerate eigen space ($\Gamma_7$ representation) under the above symmetry. In the similar way, [$\tilde\sigma_v$, $\tilde{C_3}$]$\ne$0 and [H, $\tilde\sigma_v$]= 0 together ensure the degenerate eigen state of $\psi_3$ and $\psi_4$ ($\Gamma_8$ representation). As all the points along $\Gamma$-A direction are invariant under $\tilde\sigma_v$, $\tilde{C_6}$ and $\tilde{C_3}$, these $\Gamma_7$ and $\Gamma_8$ band representations are applicable along this direction in BZ.

Furthermore, $\psi_5$ and $\psi_6$ form a degenerate state ($\Gamma_9$ representation designated by $e^{i\frac{\pi}{2}}$ and $e^{i\pi}$ eigenvalues) under the action of TRS, C$_2$ and $\sigma_d$ symmetry. As $[\tilde{C_2},\tilde\sigma_d]$= 0, we can define a new operator, $\theta$=$\tilde{C_2}$  $\tilde\sigma_d$. In spin rotation space, $\theta^2=1$. In the presence of TRS, eventually now we have, $T^{2}\theta^2=-1$. This is the local Kramer's theorem, which guaranteed the double degeneracy at every point along $\Gamma$-A direction in BZ as this direction is invariant under both $\tilde{C_2}$ and $\tilde\sigma_d$. Now, the two bands having different irreducible representations cross each other along C$_6$ axis, will form a gap less four fold degenerate Dirac nodes. For Ru, $\Gamma_{7or9}$ band intersects with another $\Gamma_8$ band and form two Dirac nodes on the C$_6$ rotation axis. The pictorial representation of bands and the formation of Dirac nodes are shown in Fig.~\ref{SM-fig1}(b).

\section{Comparison of the simulated Fermi arcs with previous ARPES results}

\begin{figure*}[h]
\centering
\includegraphics[width=\textwidth]{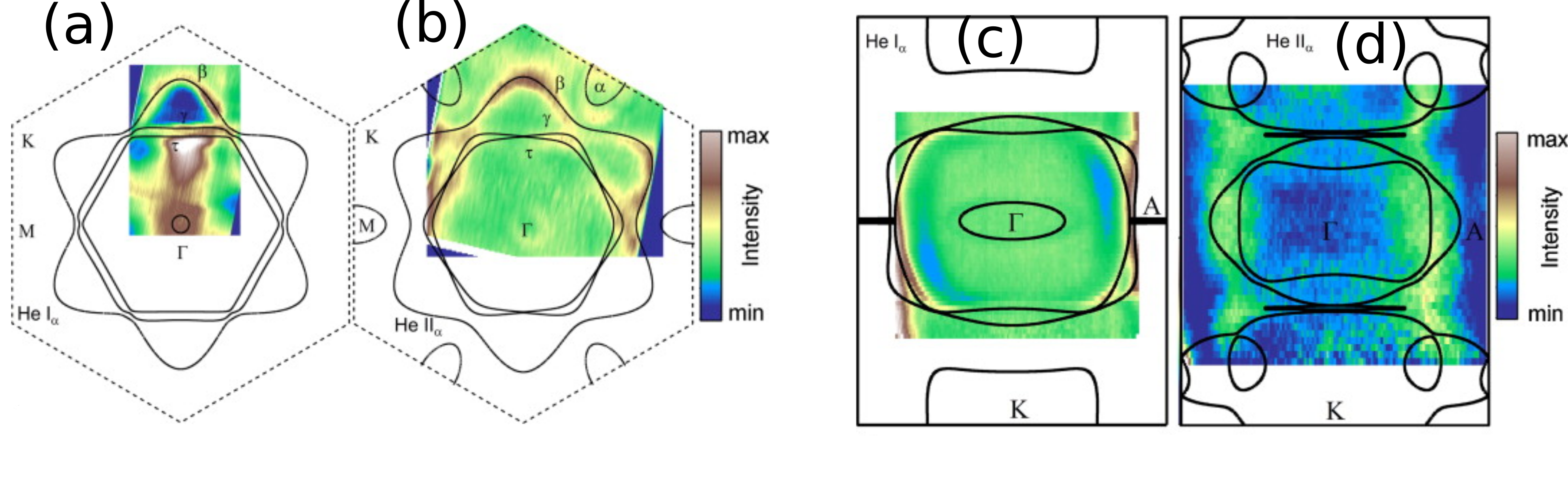}
\caption{(Color online) Experimental Fermi surfaces of Ru (001) surface (a,b) and (010) surface (c,d) measured at 33K using He I$_\alpha$  for (a,c) and He II$_\alpha$ for (b,d). The solid black lines indicate the calculated Fermi surfaces with K$_\perp$ is equal to 30$\%$ and 50 $\%$ of  $\Gamma$-A distance for (a,b) and 0$\%$ and 40 $\%$ of  $\Gamma$-A distance for (c,d) respectively. This figure is reprinted from J. Electron Spectrosc.Relat. Phenom. 191, 27-34 (2013), N. Nguyen, M. Mulazzi, and F. Reinert, with the permission from Elsevier.}
\label{SM-fig3}
\end{figure*}

In this section, we compare the topology of our simulated Fermi arcs of Ru (001) and (010) surfaces with the previously measured angle resolved photoemission spectroscopy (ARPES) results. \cite{Ru-14} The topology of our calculated Fermi surface for both surfaces are similar to that was mapped through ARPES measurements and transport experiments. \cite{Ru-14, Ru-13} Fig.~\ref{SM-fig3}(a-d) shows the Fermi arcs of Ru (001) and (010) surface measured by ARPES. The solid black lines are the calculated Fermi surfaces. The $\tau$ $\gamma$, and $\beta$ bands in Fig.~\ref{SM-fig3} are well captured in our calculations as shown in Fig.2 (b-d) in the main text. The circular/elliptical Fermi surface in (001)/(010) surfaces around the $\Gamma$ point (Fig.~\ref{SM-fig3} (a,c)) resemble with the Dirac Fermi arcs as predicted in our calculations. Please note that the calculated Fermi arcs in Fig.~\ref{SM-fig3} are corresponding to different $\Gamma$-A distance as mention in the caption of the Fig.~\ref{SM-fig3}.

\section{Formation of triple point fermion states under C$_{3v}$}

\begin{figure}[h]
\centering
\includegraphics[scale=0.5]{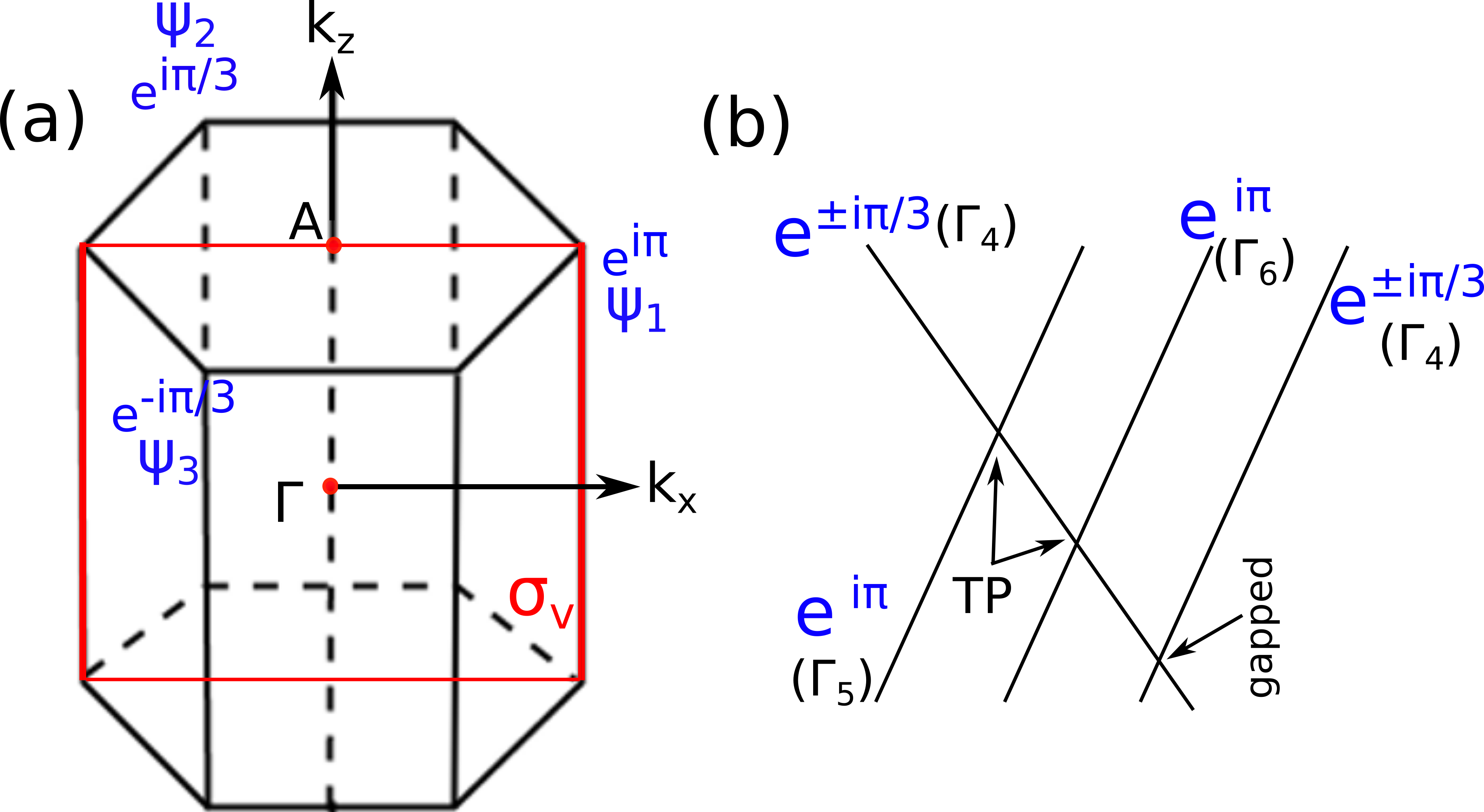}
\caption{(Color online) (a) for C$_{3v}$, the representation of rotation eigen states corresponding to their rotation eigen values, and mirror invariant planes. (b) Schematic diagram of formation of three-fold band crossings. Bands are represented by their rotation eigen values.} 
\label{SM-fig2}
\end{figure}

C$_{3v}$ have identity (E), three-fold (C$_3$) rotational symmetry about z-axis, and three vertical mirror plane ($\sigma_v$) as shown in Fig.~\ref{SM-fig2}(a). Similar to C$_{6v}$, in case of C$_{3v}, $[$\tilde\sigma_v$, $\tilde{C_3}$]$\ne$0 and [H,$\tilde\sigma_v$]= 0 ensure the degenerate eigen space of $\psi_2$ and $\psi_3$ ($\Gamma_4$ representation). Absence of $\sigma_d$ enforce to form non-degenerate representations $\Gamma_5$ and $\Gamma_6$. Any accidental band crossing $\Gamma_4$ and $\Gamma_{5or6}$ forms a triply degenerate nodal point as shown in Fig.~\ref{SM-fig2}(b).

\begin{table*}[h]
\begin{ruledtabular}
\caption{Character Table for $C_{6v}$  point group }
\label{TableS1}
\begin{tabular}{c c c c c c c c c c }
$C_{6v}$ & $E$ & $\bar{E}$ & $2C_6$ & $2\bar{C_6}$ & $2C_3$ & $2\bar{C_3}$ & $C_2$,$\bar{C_2}$ & 3$\sigma_v$,3$\bar{\sigma_v}$ & 3$\sigma_d$,3$\bar{\sigma_d}$ \\ \hline
$\Gamma_7$ & 2 & -2 & $\sqrt{3}$ & -$\sqrt{3}$ & 1 & -1 & 0 & 0 & 0 \\ \hline

$\Gamma_8$ & 2 & -2 & $-\sqrt{3}$ & $\sqrt{3}$ & 1 & -1 & 0 & 0 & 0 \\ \hline

$\Gamma_9$ & 2 & -2 & 0 & 0 & -2 & 2 & 0 & 0 & 0 \\ 

\end{tabular}
\end{ruledtabular}
\end{table*}

\begin{table*}[h!]
\begin{ruledtabular}
\caption{Character Table for $C_{3v}$  point group }
\label{TableS2}
\begin{tabular}{c c c c c c c  }
$C_{3v}$ & $E$ & $\bar{E}$ & $2C_3$ & $2\bar{C_3}$ &  3$\sigma_v$ & 3$\bar{\sigma_v}$  \\ \hline

$\Gamma_4$ & 1 & -1 & -1 & 1 & i & -i \\ \hline

$\Gamma_5$ & 1 & -1 & -1 & 1 & -i & i \\ \hline

$\Gamma_6$ & 2 & -2 & 1 & -1 & 0 & 0 \\ 

\end{tabular}
\end{ruledtabular}
\end{table*}

